\documentclass[a4paper]{article}
\usepackage{ctable,amsmath,graphicx, amsfonts,tabularx,threeparttable, float}
\newcolumntype{A}{>{\centering\arraybackslash}X}
\usepackage{cite}
\usepackage{INTERSPEECH2021}

\title{SpecMix : A Mixed Sample Data Augmentation method for Training with Time-Frequency Domain Features}

\name{Gwantae Kim$^1$, David K. Han$^2$ and Hanseok Ko$^1$}
\address{
  $^1$Korea University, South Korea\\
  $^2$Drexel University, USA}
\email{kgt1103211@korea.ac.kr, dkh42@drexel.edu, hsko@korea.ac.kr}

\begin{document}

\maketitle
\begin{abstract}
A mixed sample data augmentation strategy is proposed to enhance the performance of models on audio scene classification, sound event classification, and speech enhancement tasks. While there have been several augmentation methods shown to be effective in improving image classification performance, their efficacy toward time-frequency domain features of audio is not assured.  We propose a novel audio data augmentation approach named "Specmix" specifically designed for dealing with time-frequency domain features. The augmentation method consists of mixing two different data samples by applying time-frequency masks effective in preserving the spectral correlation of each audio sample. Our experiments on acoustic scene classification, sound event classification, and speech enhancement tasks show that the proposed Specmix improves the performance of various neural network architectures by a maximum of 2.7\%. 
\end{abstract}

\noindent\textbf{Index Terms}: Acoustic scene classification, Data augmentation, Deep neural networks, Sound event classification, Speech enhancement

\section{Introduction}
\let\thefootnote\relax\footnotetext{Corresponding Author:Hanseok Ko.}

Deep learning has shown remarkable successes on various audio processing tasks, such as speech enhancement\cite{yin2019phasen, pascual2017segan}, Automatic Speech Recognition(ASR)\cite{hannun2014deep, chan2016listen}, sound classification\cite{mcdonnell2020acoustic, piczak2015environmental, kim2020dual, lee2018time}, and speech synthesis\cite{wang2017tacotron, shen2018natural}. To further enhance their performance, many research efforts have focused on designing better network architectures for specific tasks. While improving the architectures may deliver better performances, these methods tend to overfit easily and require large amounts of training data\cite{park2019specaugment, chiu2018state}. To avoid this problem, there have been some efforts in exploring data augmentation and regularization strategies. 

For augmenting the audio dataset, there are two main approaches: time-domain waveforms and time-frequency domain features, such as spectrogram, mel-spectrogram, and mel-frequency cepstral coefficient. For the waveform data, the data augmentation strategies may include noise injection, changing pitch, changing speed, shifting time, and speed perturbation\cite{ko2015audio} of the waveform, to expand the data set without disturbing salient information therein. For the time-frequency domain features, Specaugment\cite{park2019specaugment} proposed time warping, frequency-masking, and time-masking data augmentation strategies. Although Specaugment is successfully applied to ASR, its application to other tasks has been limited\cite{park2020specaugment}. For example, in the speech enhancement task, zero-masking on the time and frequency axis tends to degrade the performance.

Since the time-frequency domain features are two dimensional and can be projected as a 2D image, data augmentation strategies, particularly of Mixed Sample Data Augmentation (MSDA) type in the computer vision domain, have been applied to the time-frequency domain features as shown in Fig. 1. Mixup\cite{zhang2017mixup, suh2020designing} blends two images of the audio features and labels by varying a random parameter $\gamma$. Its performance has been shown to be effective in the image classification tasks, however, due to the way it mixes magnitudes of spectrograms from different source components together it is difficult to disentangle them in the audio domain. Thus, the performance from the Mixup approach has been limited. Cutout\cite{devries2017improved} and Specaugment\cite{park2019specaugment} employ zero-masking to the image and spectrogram, respectively. Although these methods can be applied to images and spectrogram successfully, salient audio information can be lost due to zero-masking. Cutmix\cite{yun2019cutmix} randomly attaches a part of an image to another image. It applies a randomly generated mask for cutting a spectrogram region and pasting it randomly to another spectrogram region. While Cutmix can preserve magnitude information of $X_1$ and $X_2$, the time-frequency information taken from one image is randomly shifted to another resulting in a frequency shift.

%These methods are proposed for image classification task, but \citeshows that Mixup can also be applied to acoustic scene classification although it is not optimized to time-frequency domain feature. 

\begin{figure}
\centerline{\includegraphics[width=\columnwidth]{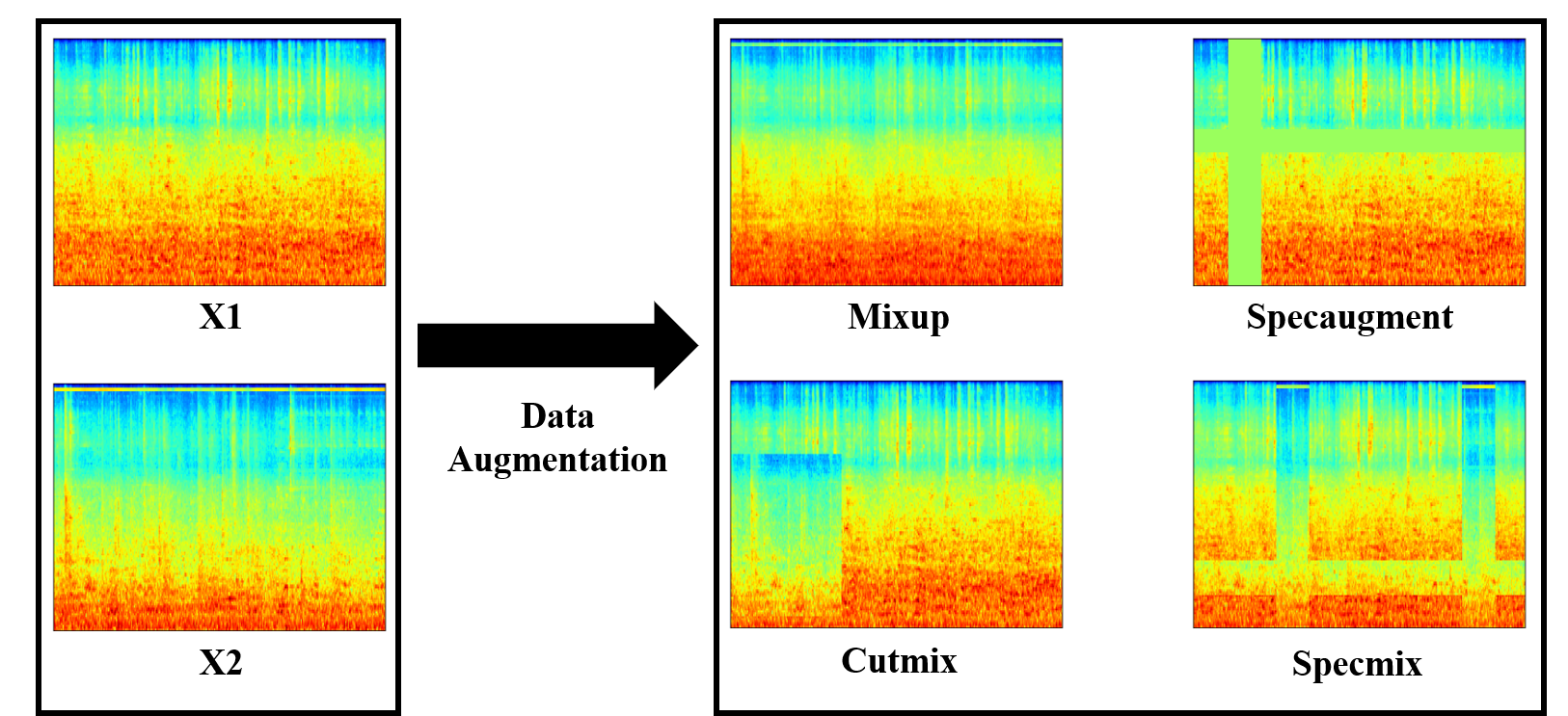}}
\caption{Overview of the data augmentation methods:Mixup, Cutmix, Specaugment and our Specmix.}
\end{figure}

%In Mixed Sample Data Augmentation(MSDA) methods, such as Mixup and Cutmix, finding an appropriate masking policy is important to preserve data distribution\cite{harris2020fmix}. However, Mixup, Cutout, and Cutmix were originally proposed for image data, their effectiveness has been limited in the time-frequency domain applications. Unlike image data, time-frequency domain features have different information distributions and correlations on time and frequency axes. 

As these methods have been adopted from image domain applications, their approach of mixing two different spectrograms has to be tailored for audio signals to preserve data distribution within a spectrogram structure for maintaining salient time-frequency correlations. Therefore, a different masking policy that preserves salient frequency features for MSDA is needed. Inspired by the previous data augmentation strategies in the speech and vision domains, we propose a novel audio data augmentation strategy, named SpecMix, for training with time-frequency domain features. The proposed method expands the idea of Cutmix, which attempted to cut-and-mix two data samples. However, their masking policy is based on image data, therefore it is not necessarily suitable for time-frequency domain features. To address this problem, we modified the masking policy tailored to time-frequency domain features. The proposed method can be integrated to ResNet, U-Net, and other state-of-the-art architectures for acoustic scene classification, sound event classification, or speech enhancement tasks.

%Fig. 1 provides an overview of Mixup, Cutmix, Specaugment and Specmix. Mixup can preserve time and frequency information of $X_1$ and $X_2$, but cannot preserve magnitude information of them because they are mixed without masking. Cutmix can preserve magnitude information of $X_1$ and $X_2$, but time and frequency information are missing because the mask is sampled on rectangular shape. Specaugment lose some information because it apply zero-masking. On the other hand, Specmix can preserve time, frequency and magnitude information of $X_1$ and $X_2$.

\section{Specmix Policy}
In this section, we describe the Specmix algorithm in detail. We aim to construct an MSDA policy that directly acts on the time-frequency domain features, which improves the generalization of the model using time-frequency domain features. 

\subsection{Algorithm on classification tasks}
Let $x\in \mathbb{R}^{F\times T\times C}$ and y denote a time-frequency domain features and its label, respectively. F denotes the number of the frequency bin, T denotes the number of the time bin and C denotes the number of time-frequency domain features. The goal of Specmix is to generate a new training sample $(\tilde{x}, \tilde{y})$ by combining two training samples $(x_A, y_A)$ and $(x_B, y_B)$. We define the combining operation as
\begin{align}
        \tilde{x} &= \mathbf{M}\odot x_A + (\mathbf{1}-\mathbf{M}) \odot x_B \\
        \tilde{y} &= \lambda y_A + (1-\lambda)y_B
\end{align}
where $\mathbf{M} \in \{0, 1\}^{F\times T}$ denotes a binary mask indicating where to drop out and fill in from two images, $\mathbf{1}$ is a binary mask filled with ones, and $\odot$ is element-wise multiplication. The combination ratio $\lambda$ between two data points is the number of pixels of $x_A$ in $\tilde{x}$.

In each training iteration, a mixed sample $(\tilde{x}, \tilde{y})$ is generated by combining two training samples selected from two mini-batches according to Equation (1) and (2).

\subsection{Algorithm on speech enhancement tasks}
Let $x\in \mathbb{R}^{F\times T\times C}$ and $z\in \mathbb{R}^{F\times T\times C}$ denote a time-frequency domain features of the noisy signal and related time-frequency domain features of the clean signal, respectively. The new training sample pair, $(\tilde{x}, \tilde{z})$, can be constructed by combining two training samples $(x_A, z_A)$ and $(x_B, z_B)$. We define the combining operation as
\begin{align}
        \tilde{x} &= \mathbf{M}\odot x_A + (\mathbf{1}-\mathbf{M}) \odot x_B \\
        \tilde{z} &= \mathbf{M}\odot z_A + (\mathbf{1}-\mathbf{M}) \odot z_B
\end{align}
where $\mathbf{M} \in \{0, 1\}^{F\times T}$ denotes a binary mask indicating where to drop out and fill in from two images, $\mathbf{1}$ is a binary mask filled with ones, and $\odot$ is element-wise multiplication. 
In each training iteration, a mixed sample pair, $(\tilde{x}, \tilde{z})$, is generated by combining two training samples selected from two mini-batches according to Equation (1) and (2).

\subsection{Masking}
Figure 2 outlines the masking process of mixing audio features from two different samples. From frequency masking, we generate frequency bands up to three different segments while the time masking does the same in generating up to three different temporal bands. The starting frequencies and time are chosen from a random process of a uniform distribution from 0 to 1 while their widths are defined by the user. These bands form the binary mask enabling the mixing of two audio spectrograms.

\begin{enumerate}
    \item Frequency masking: Up to three frequency bands can be masked depending on a random number $f_{times}$ sampled from a uniform distribution from 0 to 3 in an integer value. In each frequency band selection, the starting band frequency $f_{start}$ is selected by a random number from a uniform distribution from 0 to F. Next, the ending band frequency is calculated by the equation $f_{end} = f_{start} + \gamma F$ with $\gamma$ chosen by a user-defined parameter between 0 to 1. Repeat this process $f_{times}$ times.
    
    \item Time masking: Again, up to three time bands can be masked in the identical process utilized for the Frequency masking described above.  Again, $\gamma$, chosen by the user controls the width of the time band, and the random processes control the starting time and the number of temporal bands. 
    
\end{enumerate}

\begin{figure}
\centerline{\includegraphics[width=\columnwidth]{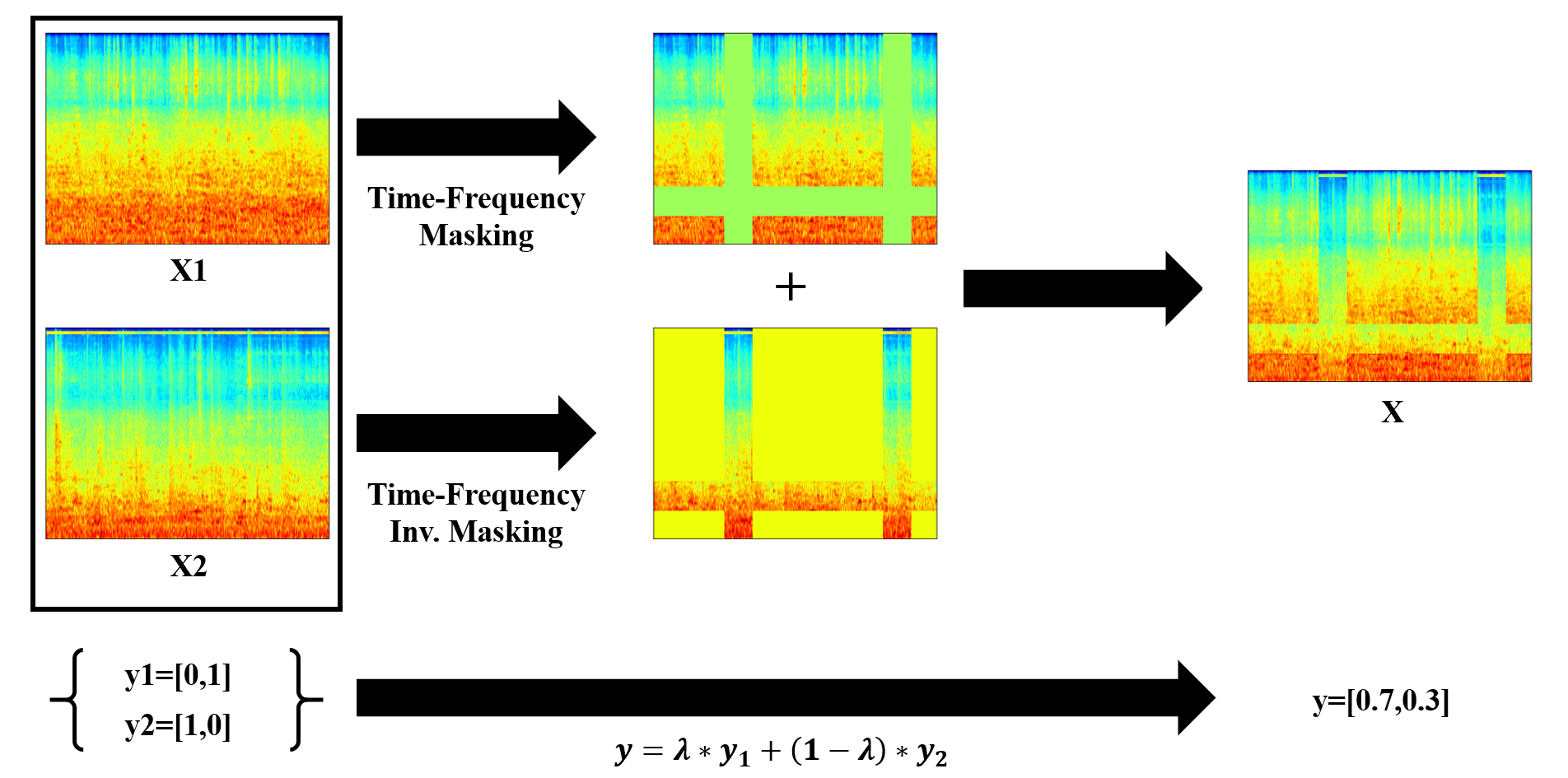}}
\caption{Augmentation policies applied to the base input, given $X_1$ and $X_2$. To generate new sample X, applying a time-frequency mask to $X_1$, inverted time-frequency mask to $X_2$, and summation of them.}
\end{figure}

\section{Models}
In this section, we introduce the preprocessing methods and models that were used to evaluate the proposed data augmentation strategy.
\subsection{Acoustic scene classification \& Sound event classification}
The input time-frequency domain features are mel-spectrogram, its delta and delta-delta features. The sampling rate is 44.1kHz, nfft is 2048 and hop length is 1024, and the number of mel filters is 128. The shape of processed input is $[B,F,T,C] = [B, 128, T, 3]$, where B denotes the batch size, F denotes the number of frequency bins, T denotes the total time bins and C denotes the total number of channels. T depends on the audio length. 

We used Resnet-101\cite{he2016deep} on acoustic scene classification and sound event classification tasks. Adam\cite{kingma2014adam} optimizer and cross-entropy loss function were used to train our model with a batch size of 32. We employed learning rate decay from 1e-3 to 1e-7. We also employed the model and training procedure proposed in \cite{mcdonnell2020acoustic}, considered state-of-the-art acoustic scene classification model, to evaluate the generalization performance of the proposed method.

\subsection{Speech enhancement}
During training, the input waveform was cut or padded to make the waveform length of 32768 samples. The input time-frequency domain feature is a spectrogram. The sampling rate is 16kHz, nfft is 512, and hop length is 256. The shape of the processed input is $[B,F,T,C] = [B, 256, T, 2]$, where B denotes the batch size, F denotes frequency bins, T denotes time bins and C denotes channels. T depends on the audio length. The first channel is the real part of the spectrogram and the second channel is the imaginary part of the spectrogram.

We build a. U-Net\cite{ronneberger2015u} style speech enhancement model illustrated in Fig. 3. The model consists of 8 encoder layers, 1 mid-level layer, 8 decoder layers, and 1 last convolution layer. To deal with arbitrary length inputs, the strides of all the layers on the time axis are not squeezed. The model predicts phase sensitive mask\cite{erdogan2015phase} and reconstructs a clean spectrogram by applying the mask to a noisy spectrogram. During training, Adam\cite{kingma2014adam} optimizer and a mean squared error loss function is used to train our model with a batch size of 6. We employ learning rate decay from 1e-2 to 1e-5.

\begin{figure}
\centerline{\includegraphics[width=\columnwidth]{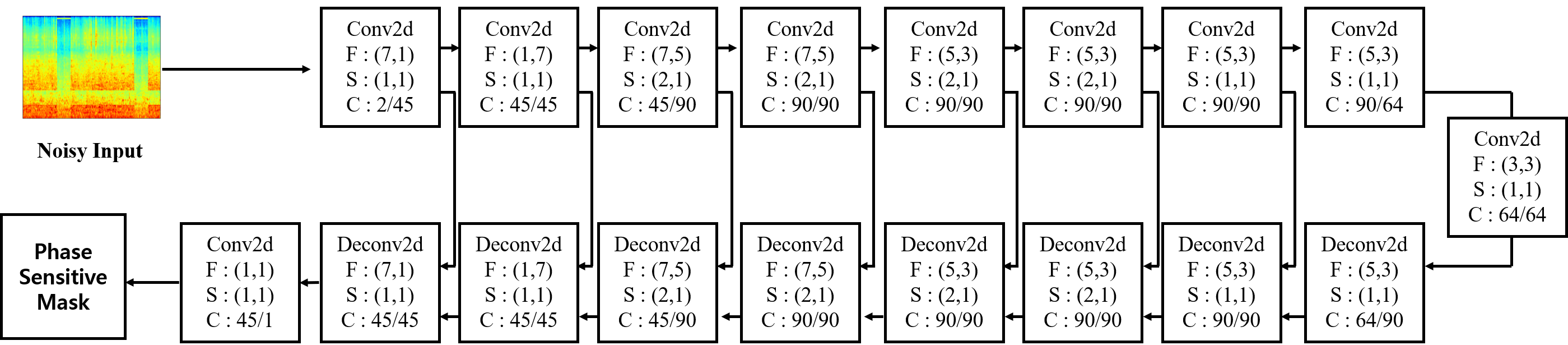}}
\caption{A U-Net model for solving speech enhancement task.}
\end{figure}

\section{Experiments}
In this section, we evaluate Specmix on acoustic scene classification, sound event classification, and speech enhancement tasks. The effects of Specmix on acoustic scene classification and sound event classification are examined first. We then compare the result with changing $\gamma$ and masking policies. We also show that Specmix can improve noise reduction performance.

\subsection{Acoustic scene classification}
We evaluate Specmix on TAU Urban Acoustic Scenes 2020 Mobile benchmark\cite{Mesaros2018_DCASE}, the dataset containing recordings from 12 European cities in 10 different acoustic scenes using 4 different recording devices. The dataset consists of 10 classes of urban acoustic scene recordings with 13965 labeled clips for training and 2970 clips for the test. The evaluation metric is accuracy.

Results with ResNet-101 model are given in Table 1. We observe that Specmix achieves the best result (62.13\% accuracy) among the considered augmentation strategies. Specmix outperforms Mixup, Cutmix and Specaugment by +3.98\%, +2.59\% and +4.45\%, respectively. Interestingly, Mixup, Cutmix, and Specaugment performed worse than with no augmentation. We believe that these augmentation strategies resulted in some information loss, as mentioned in Section 1, leading to poor performances.

\begin{table}[t]
\caption{Comparison of state-of-the-art data augmentation methods for time-frequency domain features on TAU Urban Acoustic Scenes 2020 Mobile benchmark.(Model : ResNet-101)}
\centering
    \resizebox{\linewidth}{!}{
    \begin{tabularx}{\linewidth}{|A||A||A|}
    \specialrule{.2em}{.1em}{.1em}
    \multicolumn{1}{A|}{Model : ResNet-101} & 
    \multicolumn{1}{A}{Accuracy(\%)} \\ \hline
    \multicolumn{1}{A|}{No augmentation} & 
    \multicolumn{1}{A}{59.60} \\ 
    \multicolumn{1}{A|}{Mixup\cite{zhang2017mixup}} & 
    \multicolumn{1}{A}{58.15} \\ 
    \multicolumn{1}{A|}{Cutmix\cite{yun2019cutmix}} & 
    \multicolumn{1}{A}{59.54} \\ 
    \multicolumn{1}{A|}{Specaugment\cite{park2019specaugment}} & 
    \multicolumn{1}{A}{57.68} \\
    \multicolumn{1}{A|}{Specmix $\gamma=0.3$} & 
    \multicolumn{1}{A}{\textbf{62.13}} \\
    \specialrule{.2em}{.1em}{.1em}
    \end{tabularx}}
\end{table}

\begin{figure}
\centerline{\includegraphics[width=0.8\columnwidth]{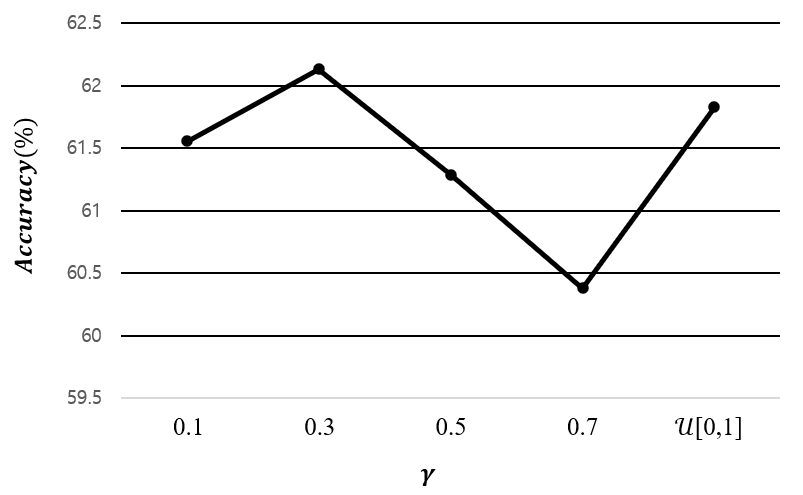}}
\caption{Impact of $\gamma$ on TAU Urban Acoustic Scenes 2020 Mobile benchmark.}
\end{figure}

We evaluate Specmix with $\gamma \in \{0.1, 0.3, 0.5, 0.7\}$ and  $\mathcal{U}[0,1]$ with ResNet-101 model. The results are given in Fig. 4. For all $\gamma$ values considered, Specmix improves the performance over the case with no augmentation. The best performance is achieved when $\gamma=0.3$.

Results with Specmix and the other augmentation methods applied to \cite{mcdonnell2020acoustic}'s model are given in Table 2. We observe that Specmix improves the performance by +0.31\%, +0.68\%, +2.53\% compared to Mixup, Cutmix, and Specaugment, respectively. We also performed an ablation study to observe the impacts of the masking strategies. The results are given in Table 3. Random masking is a masking strategy that randomly selects N pixels filled by X1 while the remaining pixels are filled by X2. Specmix(time only) only applies time masking and Specmix(freq. only) only applies frequency masking.
The results show that both time masking and frequency masking improve performance.

\begin{table}[t]
\caption{Comparison of state-of-the-art data augmentation methods for time-frequency domain features on TAU Urban Acoustic Scenes 2020 Mobile benchmark.(Model : \cite{mcdonnell2020acoustic})}
\centering
    \resizebox{\linewidth}{!}{
    \begin{tabularx}{\linewidth}{|A||A||A|}
    \specialrule{.2em}{.1em}{.1em}
    \multicolumn{1}{A|}{Model : \cite{mcdonnell2020acoustic}} & 
    \multicolumn{1}{A}{Accuracy(\%)} \\ \hline
    \multicolumn{1}{A|}{No augmentation} & 
    \multicolumn{1}{A}{68.90} \\ 
    \multicolumn{1}{A|}{Mixup\cite{zhang2017mixup}} & 
    \multicolumn{1}{A}{71.29} \\ 
    \multicolumn{1}{A|}{Cutmix\cite{yun2019cutmix}} & 
    \multicolumn{1}{A}{70.92} \\ 
    \multicolumn{1}{A|}{Specaugment\cite{park2019specaugment}} & 
    \multicolumn{1}{A}{69.07} \\
    \multicolumn{1}{A|}{Specmix $\gamma=\mathcal{U}[0,1]$} & 
    \multicolumn{1}{A}{\textbf{71.60}} \\
    \specialrule{.2em}{.1em}{.1em}
    \end{tabularx}}
\end{table}

\begin{table}[t]
\caption{Impact of masking strategies on TAU Urban Acoustic Scenes 2020 Mobile benchmark.(Model : \cite{mcdonnell2020acoustic})}
\centering
    \resizebox{\linewidth}{!}{
    \begin{tabularx}{\linewidth}{|A||A||A|}
    \specialrule{.2em}{.1em}{.1em}
    \multicolumn{1}{A|}{Model : \cite{mcdonnell2020acoustic}} & 
    \multicolumn{1}{A}{Accuracy(\%)} \\ \hline
    \multicolumn{1}{A|}{Random masking} & 
    \multicolumn{1}{A}{70.05} \\
    \multicolumn{1}{A|}{Specmix(time only)} & 
    \multicolumn{1}{A}{70.42} \\
    \multicolumn{1}{A|}{Specmix(freq. only)} & 
    \multicolumn{1}{A}{70.52} \\
    \multicolumn{1}{A|}{Specmix} & 
    \multicolumn{1}{A}{\textbf{70.79}} \\
    \specialrule{.2em}{.1em}{.1em}
    \end{tabularx}}
\end{table}

\subsection{Sound event classification}
We evaluate on SECL\_UMONS benchmark\cite{brousmiche2020secl}. SECL\_UMONS is a real sound recording dataset for simultaneous classification and localization of sound events. The data samples in the SECL\_UMONS are recording in two indoor room environments. The room condition of the first one is RT60 of 0.7s with the room dimensions of $7.8\times3.6\times2.45$[m] and the second one is RT60 of 0.9s with room dimensions of $9.4\times7.5\times4.85$[m]. The dataset consists of 11 indoor event classes, such as \emph{chair\_movement}, \emph{cup\_drop\_off}, \emph{furniture\_drawer}, \emph{hand\_clap}, \emph{keyboard}, \emph{knock}, \emph{phone\_ring}, \emph{radio}, \emph{speaker}, \emph{step}, and \emph{whistle}. The dataset contains 2178 sequences for the training set and 484 sequences for the validation set. We used single-channel waveform for our comparison. Evaluate metric is accuracy.
Results with the ResNet-101 model are given in Table 4. Specmix achieves 97.107\% accuracy on SECL\_UMONS, +1.04\% higher than the no augmentation case.

\begin{table}[t]
\caption{Comparison of state-of-the-art data augmentation methods for time-frequency domain features on SECL\_UMONS benchmark.(Model : ResNet-101)}
\centering
    \resizebox{\linewidth}{!}{
    \begin{tabularx}{\linewidth}{|A||A||A|}
    \specialrule{.2em}{.1em}{.1em}
    \multicolumn{1}{A|}{Model : ResNet-101} & 
    \multicolumn{1}{A}{Accuracy(\%)} \\ \hline
    \multicolumn{1}{A|}{No augmentation} & 
    \multicolumn{1}{A}{96.07} \\ 
    \multicolumn{1}{A|}{Mixup\cite{zhang2017mixup}} & 
    \multicolumn{1}{A}{95.87} \\ 
    \multicolumn{1}{A|}{Cutmix\cite{yun2019cutmix}} & 
    \multicolumn{1}{A}{95.66} \\ 
    \multicolumn{1}{A|}{Specaugment\cite{park2019specaugment}} & 
    \multicolumn{1}{A}{96.90} \\
    \multicolumn{1}{A|}{Specmix $\gamma=\mathcal{U}[0,1]$} & 
    \multicolumn{1}{A}{\textbf{97.11}} \\
    \specialrule{.2em}{.1em}{.1em}
    \end{tabularx}}
\end{table}

\subsection{Speech enhancement}
We evaluated Specmix on Voicebank + Diverse Environments Multichannel Acoustic Noise Database(DEMAND) benchmark, which is proposed by \cite{valentini2016investigating}. Noisy and clean speech recordings were provided from the DEMAND\cite{thiemann2013diverse} and the Voice Bank corpus\cite{veaux2013voice}, respectively with each recorded with the sampling rate of 48kHz. A total of 40 different noise conditions are considered in the training set and 20 different conditions are considered in the test set. Finally, the training and test set contained 11572 and 824 noisy-clean speech pairs, respectively. Note that the speaker and noise classes were uniquely selected for the training and test sets. Evaluation metrics are Perceptual Evaluation of Speech Quality(PESQ), mean opinion score predictor of signal distortion(CSIG), background noise intrusiveness(CBAK), overall signal quality(COVL), and Segmental Signal to Noise Ratio(SSNR).

Results with the U-Net model are given in Table 5, where NA denotes No Augmentation, MU denotes Mixup, CM denotes Cutmix, and SA denotes SpecAugment. Mixup and Specaugment have failed to improve the speech enhancement performance over the no augmentation. However, Cutmix improves the performance of speech enhancement, and Specmix achieves the best results compared to the other data augmentation strategies.

We evaluate Specmix with $\gamma \in \{0.1, 0.3, 0.5, \mathcal{U}[0,1]\}$ with U-Net model. The results are given in Fig. 5. Although it is not shown, all the cases with  $\gamma$ values utilized in Specmix improved performance over the case with no augmentation. The best performance is achieved when $\gamma=0.3$.
\begin{table}[t]
\caption{Comparison of state-of-the-art data augmentation methods for time-frequency domain features on Voice bank + DEMAND benchmark.(Model : U-Net).}
\centering
    \resizebox{\linewidth}{!}{
    \begin{tabularx}{\linewidth}{|A||A||A||A||A||A|}
    \specialrule{.2em}{.1em}{.1em}
    \multicolumn{1}{A|}{} & 
    \multicolumn{1}{A}{PESQ} & 
    \multicolumn{1}{A}{CSIG} & 
    \multicolumn{1}{A}{CBAK} & 
    \multicolumn{1}{A}{COVL} & 
    \multicolumn{1}{A}{SSNR} \\ \hline
    \multicolumn{1}{A|}{Noisy} & 
    \multicolumn{1}{A}{1.97} & 
    \multicolumn{1}{A}{3.35} & 
    \multicolumn{1}{A}{2.44} & 
    \multicolumn{1}{A}{2.63} & 
    \multicolumn{1}{A}{1.67} \\ 
    \multicolumn{1}{A|}{NA} & 
    \multicolumn{1}{A}{2.50} & 
    \multicolumn{1}{A}{3.44} & 
    \multicolumn{1}{A}{3.18} & 
    \multicolumn{1}{A}{2.95} & 
    \multicolumn{1}{A}{9.26} \\ 
    \multicolumn{1}{A|}{MU} & 
    \multicolumn{1}{A}{2.44} & 
    \multicolumn{1}{A}{3.39} & 
    \multicolumn{1}{A}{3.16} & 
    \multicolumn{1}{A}{2.89} & 
    \multicolumn{1}{A}{9.43} \\ 
    \multicolumn{1}{A|}{CM} & 
    \multicolumn{1}{A}{2.52} & 
    \multicolumn{1}{A}{3.50} & 
    \multicolumn{1}{A}{3.22} & 
    \multicolumn{1}{A}{2.99} & 
    \multicolumn{1}{A}{9.48} \\ 
    \multicolumn{1}{A|}{SA} & 
    \multicolumn{1}{A}{2.41} & 
    \multicolumn{1}{A}{3.48} & 
    \multicolumn{1}{A}{3.16} & 
    \multicolumn{1}{A}{2.93} & 
    \multicolumn{1}{A}{9.35} \\
    \multicolumn{1}{A|}{Specmix} & 
    \multicolumn{1}{A}{\textbf{2.54}} & 
    \multicolumn{1}{A}{\textbf{3.60}} & 
    \multicolumn{1}{A}{\textbf{3.24}} & 
    \multicolumn{1}{A}{\textbf{3.05}} & 
    \multicolumn{1}{A}{\textbf{9.57}} \\
    \specialrule{.2em}{.1em}{.1em}
    \end{tabularx}}
\end{table}

\begin{figure}
\centerline{\includegraphics[width=0.9\columnwidth]{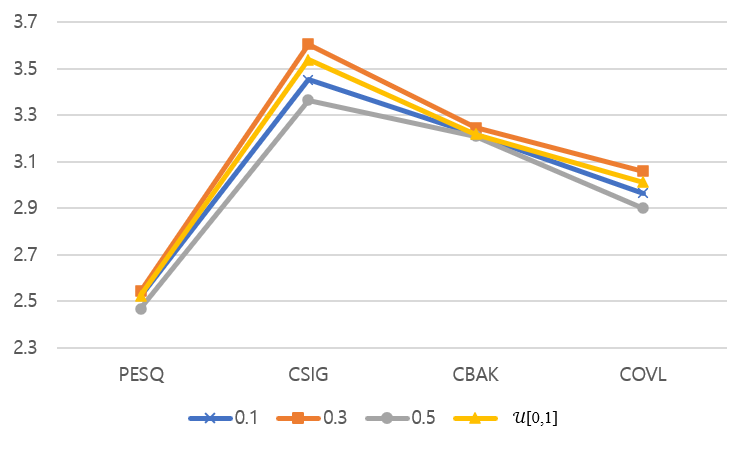}}
\caption{Impact of $\gamma$ on Voice Bank + DEMAND benchmark.}
\end{figure}

\section{Conclusion}
In this paper, we proposed a data augmentation strategy, named Specmix, for training with time-frequency domain features. 
While there are augmentation strategies for mixing two different audio sources, our proposed method was shown to preserve spectral information throughout the augmentation process.
%First, time-frequency mask is selected with proposed masking policies. Second, applying mask to first sample and inverse mask to second sample. Finally, summing two masked samples and generate new sample. 
The proposed Specmix was shown easy to be incorporated into existing training pipelines and it has a low computational cost. From several evaluations of the proposed augmentation, the method improved the performance over various models on audio scene classification, sound event classification, and speech enhancement tasks. We expect that Specmix can be applied to various machine learning tasks using time-frequency domain features. 

\section{Acknowledge}
This material is based upon work supported by the Air Force Office of Scientific Research under award number FA2386-19-1-4001.

\bibliographystyle{IEEEtran}
\bibliography{ref}

\end{document}